# A 1D Optomechanical crystal with a complete phononic band gap.

J. Gomis-Bresco[1*], D. Navarro-Urrios[1*], M. Oudich[2], S. El-Jallal[2,3], A. Griol[4], D. Puerto[4], E. Chavez[1,5], Y. Pennec[2], B. Djafari-Rouhani[2], F. Alzina[1], A. Martínez[4] and C. M. Sotomayor Torres[1, 6].

[1] *ICN2 - Institut Catala de Nanociencia i Nanotecnologia, Campus UAB, 08193 Bellaterra (Barcelona), Spain*
[2] *IEMN, Universite de Lille 1, Villeneuve d'Ascq, France*
[3] *PRILM, Université Moulay Ismail, Faculté des sciences, Meknès, Maroc*
[4] *Nanophotonics Technology Center, Universitat Politècnica de València, Valencia, Spain*
[5] *Dept. of Physics, Universitat Autònoma de Barcelona, 08193 Bellaterra (Barcelona), Spain.*
[6] *ICREA - Institucio Catalana de Recerca i Estudis Avançats, 08010 Barcelona, Spain*

*Corresponding author: jordi.gomis@icn.cat*
* *These authors contributed equally to this manuscript.*

Abstract. **Recent years have witnessed the boom of cavity optomechanics, which exploits the confinement and coupling of optical and mechanical waves at the nanoscale[1,2]. Amongst their physical implementations[3], optomechanical (OM) crystals [4,5] built on semiconductor slabs enable the integration and manipulation of multiple OM elements in a single chip and provide GHz phonons suitable for coherent phonon manipulation[6,7]. Different demonstrations of coupling of infrared photons and GHz phonons in cavities created by inserting defects on OM crystals have been performed[4,8–11]. However, the considered structures do not show a complete phononic bandgap, which should allow longer dephasing time, since acoustic leakage is minimized. We demonstrate the excitation of acoustic modes in a 1D OM crystal properly designed to display a full phononic bandgap for acoustic modes at 4 GHz. The modes inside the complete bandgap are designed to have mechanical Q factors above $10^8$ and invariant to fabrication imperfections.**

Optomechanical (OM) coupling, the direct interaction of electromagnetic radiation and mechanical vibrations of matter, can be greatly enhanced by confining electromagnetic radiation in a cavity. Since the establishment of cavity optomechanics[1,12] the transduction of phonons through light has led to a wide variety of applications in sensing[5,13,14] and communications[15–18]. Moreover, the manipulation of the mechanical degrees of freedom with light, with the demonstrations of energy transfer from photons to phonons (amplification[19]) and from phonons to photons (sideband resolved OM cooling[20,21]) has paved the way towards the coherent quantum control of a mechanical oscillator[6,22].

Recent experimental demonstration of cooling down to the ground state of a mechanical resonator[8,23] with less than a single confined phonon in average, and coherent effects like optomechanically induced transparency[9,24] and coherent coupling in the well resolved sideband regime (OM normal mode splitting [3]) confirm OM cavities as ideal building blocks for quantum computation using phonons. But the conditions that a cavity OM system, which range from the macroscopic to the atomic domain and make use of radiation covering the microwave to the visible domain[25], must fulfil to allow phonon coherent manipulation are strongly restrictive. The OM vacuum coupling (g) strength is one of the key factors, with the highest reports giving $g/2\pi$ rates in the MHz range [5,26–28] and reports of GHz in extended systems[29]. This g factor multiplied by the photon lifetime in the cavity ($\tau_{cav}$) gives an important figure of merit ($g/\kappa$ where $\kappa$ is the optical cavity decay rate, $\kappa=1/\tau_{cav}$), which can be related with the interaction strength per photon inside the cavity. Ground state cooling, that is, average phonon population below 1 at the ground state, requires that the OM system is within the side band resolved limit. In other words, the lifetime in the cavity must exceed the mechanical oscillation period ($T_{mec} < \tau_{cav}$).

In addition, it is important to preserve the phonon mode from the interaction with thermally populated acoustic phonons that dephase the system. High frequency phononic modes ($\Omega_{mec}$ in the GHz) are less populated for the same temperature ($n_{th} \approx k_B T_{th}/\hbar\Omega_{\mathrm{mec}}$), and can be cooled down to the ground state even without OM assisted cooling[7]. The quality of the mechanical cavity ($Q_{mec}$), a signature of the level of isolation reached in the phonon confinement, can be compensated by the intrinsic weight of the phonon frequency, so $Q_{mec}\cdot \Omega_{mec}/2\pi$ is considered a good figure of merit for quantum coherent phonon manipulation. Goryachev et al[30] reached with a microwave system $7.8\cdot10^{16}$ Hz whereas Chan et al[8] reached $3.9\cdot10^{14}$ Hz with an OM crystal.

OM crystals[4] are simultaneously photonic and phononic crystals that confine in the same structure photons and phonons, so that when engineered properly they can lead to strong photon-phonon interaction. A practical advantage remarks them over the rest of OM systems: ease of integration in chip platforms allows the design of multiple OM elements as circuits. But they have also an edge on the fundamental physical limits attainable in phonon isolation, as their typical phonon frequency lies in the high GHz range and phononic band gaps can be design to prevent phonon propagation at certain frequency ranges. A complete phononic band gap is defined by the absence of any phononic band in a given frequency range, meanwhile, a pseudo-band gap is defined by the absence of bands of a particular symmetry in the frequency range of definition, even if there are still bands of other symmetries at that frequencies. In Chang et al [8,27] , the Q mechanical factor of the confined phononic mode at cryogenic temperatures is limited by unwanted fabrication imperfections that break the perfect symmetry of the ideal (as designed) structure and allow coupling among different symmetry phononic propagative bands inside the pseudo-bandgaps. This loss is partly mitigated by surrounding the structure with a phononic radiation shell (a 2D phononic crystal with a complete band gap). By means of that approach, the energy coupled to the unwanted propagative modes of other symmetries does not leak out to the substrate, but is indeed lost by the confined phononic mode and spread around the full nanobeam. This leaking mechanism could be completely avoided by using a structure with a complete phononic band gap, but in practice the design of a 1D cavity that is at the same time photonic and phononic is even more restrictive and has not been accomplished so far. Limiting the losses would increase the dephasing time and allow multiple coherent phonon operations in a quantum computing scheme based in phonons and controlled/read out by light. As shown in a really recent work by Safavi-Naeini et al[31], the use of a full phononic bandgap is essential to reduce the loss channels in a 2D optomechanical crystal, in that case they are capable of limiting the coupling of their confined phonon mode to a single band in the Gamma-X direction of the waveguide used to create the defect.

In this paper we present a silicon 1D OM crystal built up so that it displays a dual absolute band gap for both phonons and photons[32], as proposed by Maldovan and Thomas in ref 33, what has been named a phoXonic crystal. Figure 1a sketches the top view of the unit cell of the proposed OM crystal, a 220 nm thick nanobeam. The key advantage of this cell over previous approaches is that optical properties are mainly determined by the inner beam width and the hole size and spacing whilst the mechanical properties are specially affected by the stub width and size. Such approach uncouples effectively the design of the optical and the mechanical cavities which allows achieving a full 1D phononic band gap for propagative modes and a photonic bandgap for TE-like (even parity) optical modes.

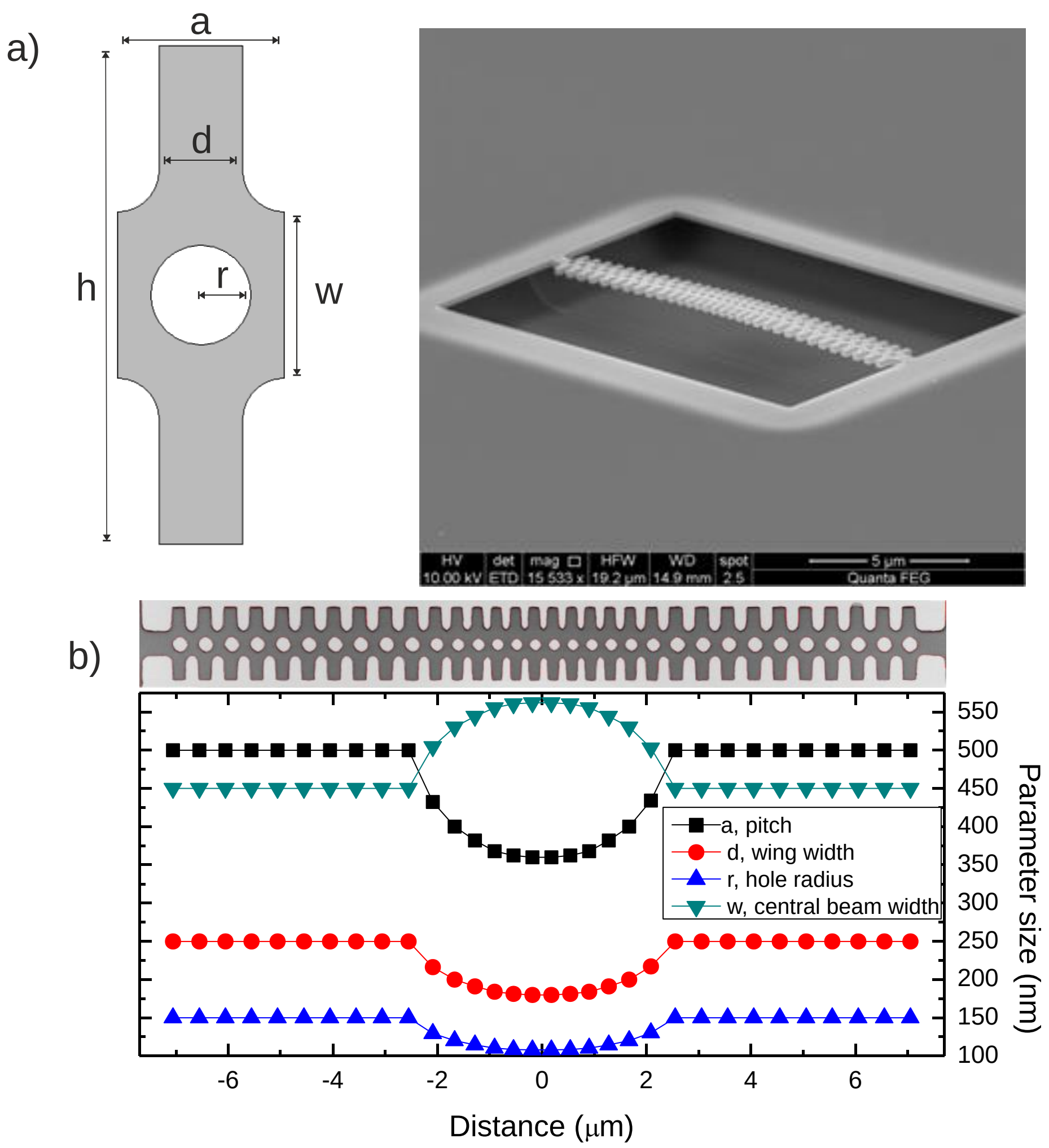

**Figure 1. Description of the 1D OM crystal with a full phononic bandgap. a), Ideal unit cell with the parameters that define it, adapted from reference[32] and tilted SEM image of typical fabricated device. b) Parameter variation to build up the OM defect crystal cavity. On top of the graph we show a SEM image of the fabricated OM cavity with dimensions corresponding one to one to the ordinate axis. The red contour of the SEM image corresponds to the profile used in the FEM calculations to model the real structure.**

The cavity is built by varying in a parabolic way towards the centre of the structure the cell width (a, pitch), hole radius (r) and stub width (d) the same percentage (Percentage of Reduction (PR)), meanwhile the beam width (w) is increased towards the centre. In the OM cavity presented in Figure 1 b), PR is 72%. These parameters and others (hole to ellipse, stub shape...) could be varied in an independent way at the expense of a numerically complex optimization process[27], what gives a taste of the room for improvement in the OM coupling still to explore. On top of the graph, we present a Scanning Electron Microscope (SEM) image of the OM cavity discussed onwards. The fabrication imperfections are taken into account in the simulations along the paper (see supplementary information). We draw in red the contour of the structure obtained rendering the SEM image. This contour is used to calculate the optical and mechanical properties of the structure using finite element method (FEM) in commercial software Comsol. We use ten mirror cells at every side of the cavity, composed by twelve cells, and centred in a beam connector.

The photonic band diagram (figure 2a) of the OM cavity's mirror cells presents a band gap for TE modes. The variation of the cell towards the centre of the OM cavity creates an effective confining potential by pulling up the modes of the second and third TE band up from the X point. The tapering allows us achieving strongly confined high Q photonic modes at telecom wavelengths. We place a tapered fibre loop (see supplementary information) in the vicinity of the device standing partially on the frame around the substrate. In all the experiments, light is coupled in and out of the device from the tapered fibre. Experimental results for the fabricated structure (figure 2b) show an optical Q factor over $1{\cdot}10^4$ for the mode at 1529 nm.  We identified six confined optical modes in the wavelength range between 1470 to 1600 nm by comparing the simulations and the experimental results.

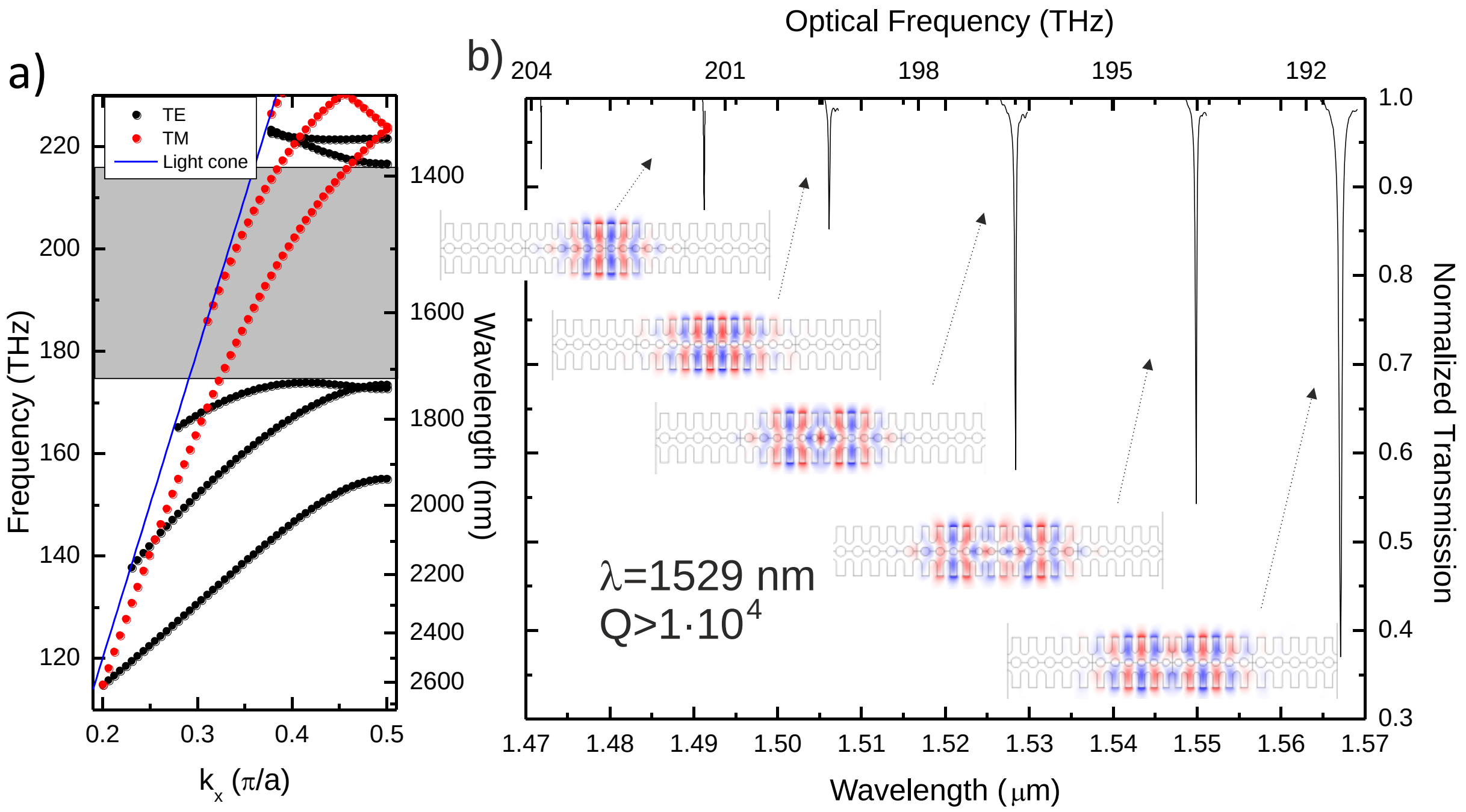

**Figure 2 Photonic properties of the OM nanobeam crystal  a) Photonic band diagram of an idealized mirror cell. In black TE optical modes, in grey the pseudo band-gap for the TE mode. b) Experimental optical spectroscopy curve of the device 1, FEM simulations of an idealized device shown as guidance**

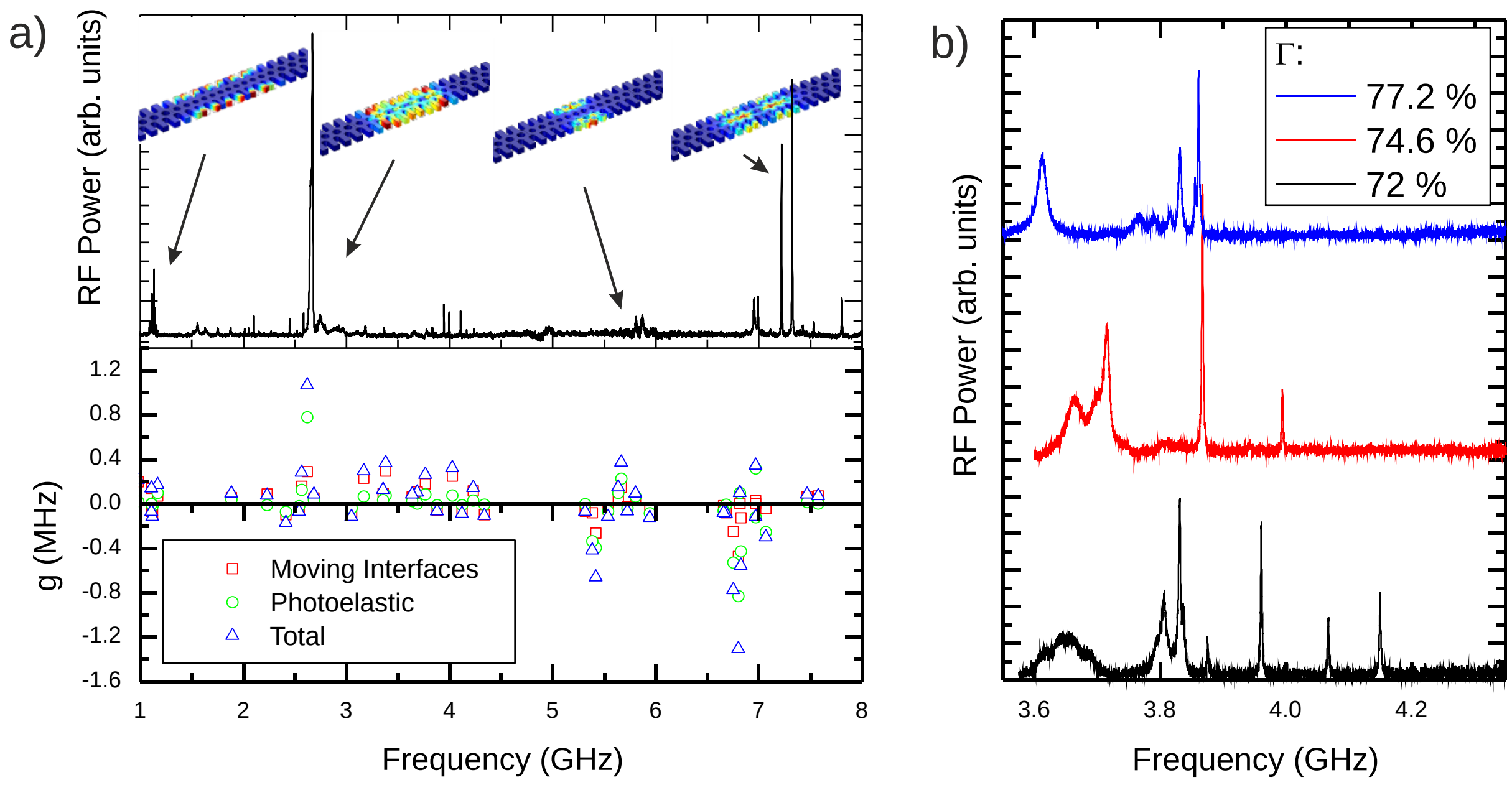

**Figure 3 Phononic properties of the OM nanobeam crystal a) Top, RF Spectrum of the phononic modes transduced by the device. Five different families are identified (four plotted, the last one can be observed in figure 4) and FEM simulations of an idealized (as designed) structure are shown. Bottom, vacuum OM coupling (g) calculations corresponding to the fabricated OM cavity of figure 1 and 2 (limited to the 50 phonon modes with higher g). We distinguish between contributions of moving interfaces (red hollow square) and photoelastic (green hollow circle). The total OM coupling is the addition of both contributions (blue hollow triangle). b) Experimental RF spectra of a different series of OM cavities where the deepness of the defect (1-PR) is increased. The confined modes shift towards higher frequency entering into the complete band gap.**

By detuning the laser to the blue at the maximum amplitude slope of the resonance at 1529 nm we are capable of transducing the mechanical vibration confined in the cavity (figure 3 a) top). We should expect to transduce only modes symmetric to the planes normal to the nanobeam (symmetric/symmetric) but fabrication imperfections break and mix the symmetries. We identify five different phononic mode families by comparing FEM simulations and the RF spectrum, exemplary displacement profiles are shown and linked with arrows to the corresponding experimental transduced modes. We confirm the grouping of the phononic modes observed by changing the minimum reduction percentage of the parabolic defect (hundreds of devices fabricated, simulated and characterized) and the optical mode used to transduce them. We achieve amplification of phononic modes from families at 5.5 and 7 GHz by measuring in coupling conditions well in the resolved band limit (supplementary information).

In figure 3 a) , we plot the vacuum OM coupling rate split in moving interfaces and photoelastic contributions for the modes confined in the cavity. The maximum absolutes rates calculate are beyond the MHz for the modes at 2.61 GHz (g=1.06 MHz) and 6.8 GHz (g=-1.3 MHz), above previous reports in OM crystal cavities and confirmed experimentally for some of the modes (supplementary information). Moreover, in most of the cases, the contributions of the moving interface and the photoelastic effects add, what is a remarkable particularity of the geometry proposed that makes it extremely interesting. The optimization process becomes more promising in this case, as up to now, the maximum OM coupling rates are obtained by maximizing the dominance of one of these effects over the other and not both of them simultaneously.

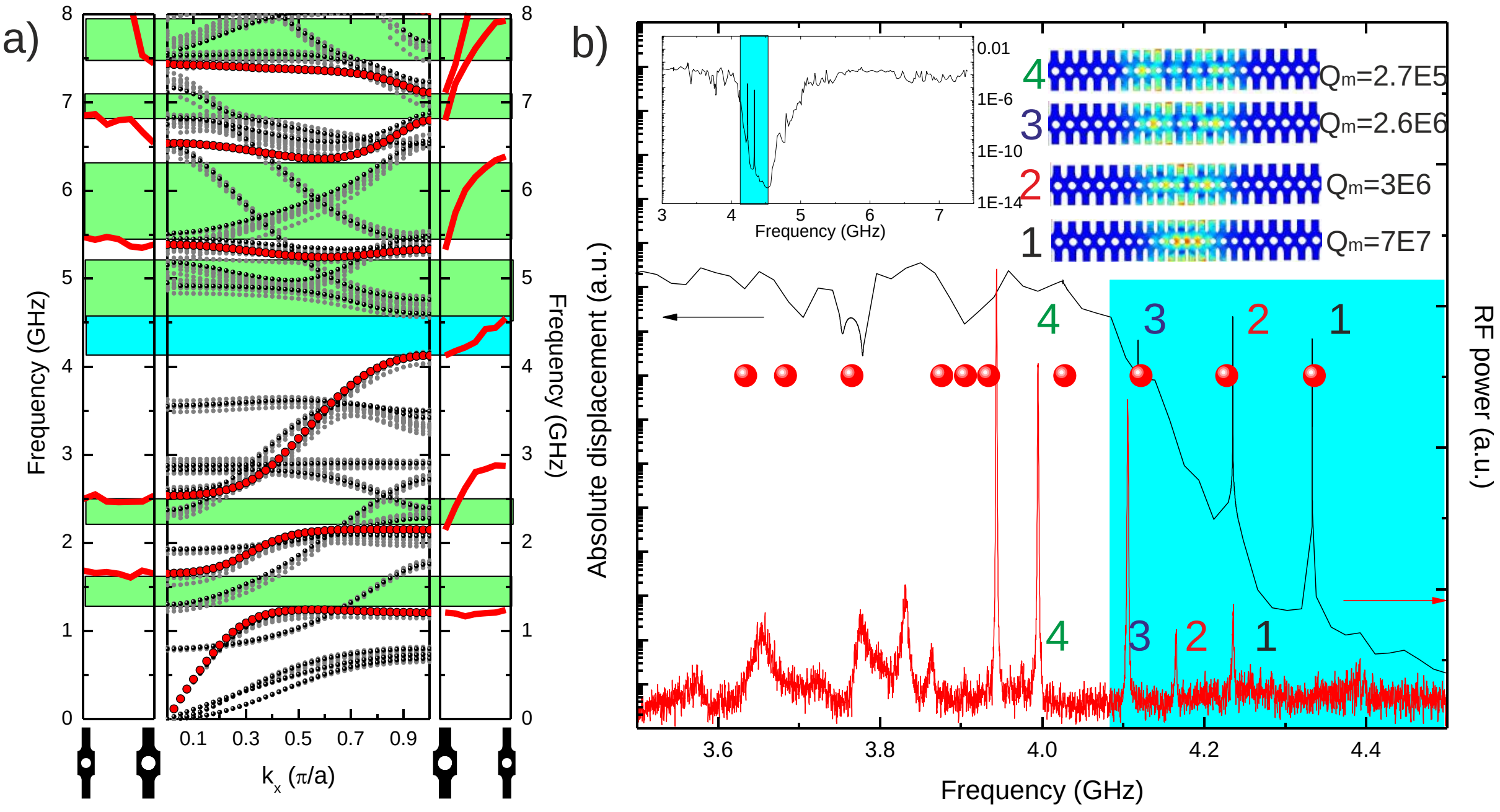

**Figure 4 Phononic bandgap a), Phononic band diagram of the mirror cells, in the central panel. In red the symmetric/symmetric bands of an idealized (as designed) mirror cell. In black the bands corresponding to combinations of other symmetries. In grey and behind the red and black curves, the simulated bands for all mirror cells as fabricated, the dispersion induced by the fabrication imperfections does not close the complete bandgap (colored light blue region) or the pseudo-bandgaps (green). The lateral panels show the symmetric-symmetric confinement potentials created for the Γ ($k_x$=0 left panel) and the X ($k_x$=π/a right panel) symmetry points as the unit cell is varied towards the centre, including fabrication imperfections. b), Comparison between the transmission simulated using the red contour of figure 1 and the RF spectroscopy measurement shown in figure 3 zoomed around the full bandgap. The FEM simulations of the confined modes displacement are shown and numbered in correspondence to the peaks of the simulations (black) and the measurements (red). The value of the mechanical Q factor is shown next to them. Phonon modes 1 to 3 lay in a full bandgap (light blue coloured region). We mark the frequency of the confined modes found in the simulation for that range with red spheres as guidance. Inset, mechanical transmission simulation, with the bandgap identified.**

The origin of the families observed can be linked directly to the phononic bands we used to create them. In figure 4 a), we plot in the central panel the phononic band diagram for an idealized (without fabrication imperfections) (black dots) and all real mirror cells (grey dots). The red dots in the same panel draw the symmetric/symmetric bands that should theoretically lead to OM coupling. At the side panels we plot the confinement potentials created by the modification of the cells towards the centre. These confinement potentials correspond perfectly with the confined modes simulated and measured in figure 3. The modes around 4 and 5.5 and above 7 GHz are drawn from the X point ($k_x$=π/a), the origin of the stronger modes at 1 and 2.5 GHz is the Γ point ($k_x$=0). The dispersion created by unwanted fabrication imperfections breaks the symmetry and allows mode coupling between modes of different symmetries, breaking the effectiveness of the mode isolation in pseudo-bandgaps (green rectangles). Nevertheless, the modes inside the complete bandgap (light blue) should be immune to such effect, as simulations (FEM simulations of figure 4 b)) show. The mechanical Q factor increases up to $7 \cdot 10^7$ for the mode placed deeper in the bandgap, even if the fabrication imperfections limit it strongly. The fabricated device presents attenuation of the transmitted signals of at least eight orders of magnitude for the zone surrounding the bandgap as simulated in the inset of figure 4 b). A fair agreement between simulations and measurements allows us to label the measured modes and identify three of them as confined inside the complete bandgap. In figure 3 b) we show how by increasing the deepness of the defect in the OM cavity (1-PR), the confined modes around the low frequency band edge of the complete bandgap can be displaced into it by increasing their frequency.

Nevertheless, the mechanical Q factor ($Q_{mec}=\Omega_{mec}\cdot\tau_{phon}$ , where $\tau_{phon}$ is the phonon decay time) measured at room temperature at atmospheric pressure is limited to values around 2000 by intrinsic phonon scattering mechanisms, like thermo-elastic decay or Akhieser. Size effects in nanostructures change the phonon lifetime compared to bulk[34] by modifying the intrinsic scattering, i.e. the thermal conductivity is strongly reduced due to diffusive boundary scattering[35]. In the regime of high frequency, extrinsic boundary scattering can become the dominant loss mechanism, when the wavelength of the phonon is comparable to the characteristic dimension of the surface roughness[34].

In summary, we designed, fabricated and measured a high Q photonic structure capable of transducing confined phononic modes inside an absolute phononic band-gap. The confined phonons have an OM coupling ranging from the KHz to the MHz range with contributions from moving interfaces and the photoelastic effect that add constructively for many of them. Using such a cavity, we demonstrate transduction of four other family modes, confined in pseudo-gaps, up to 8 GHz. Coupling light via a tapered fibre we reach the resolved side-band limit an self-amplification for various of these families. As discussed, the mechanical modes confined in the full band gap are in potential a new platform for OM phonon coherent manipulation insensitive to fabrication deviations. Therefore, OM structures displaying a phononic bandgap could be disruptive in the field of cavity optomechanics.

Acknowledgements: This work was supported by the European Commission Seventh Framework Programs (FP7) under the FET-Open project TAILPHOX N◦ 233883. J.G-B, D.N-U, E.C, F.A and C.M.S-T acknowledge financial support from the Spanish projects ACPHIN (ref. FIS2009-10150) and TAPHOR (MAT2012-31392). J.G-B and D. P. acknowledges funding from the Spanish government though the Juan de la Cierva program, D.N-U. acknowledges funding from the Catalan government through the Beatriu de Pinos program. We thank Juan Sierra for his valuable technical advice.

Contributions: M.O., J.G-B, D.N, Y.P, B.D and A.M designed the structure. M.O., S.E and J.G-B performed the simulations. A.G fabricated the samples, D.N, J.G-B, F.A and D.P performed the experiments and analysed the data. All authors contributed to the writing and discussion of the manuscript.

## Bibliography.

# Supplementary information: A 1D Optomechanical crystal with a complete phononic band gap.

J. Gomis-Bresco[1*], D. Navarro-Urrios[1*], M. Oudich[2], S. El-Jallal[2,3], A. Griol[4], D. Puerto[4], E. Chavez[1], Y. Pennec[2], B. Djafari-Rouhani[2], F. Alzina[1], A. Martínez[4] and C. M. Sotomayor Torres[1, 5, 6].

[1] *ICN2 - Institut Catala de Nanociencia i Nanotecnologia, Campus UAB, 08193 Bellaterra (Barcelona), Spain*
[2]*IEMN, Universite de Lille 1, Villeneuve d'Ascq, France*
[3] *PRILM, Université Moulay Ismail, Faculté des sciences, Meknès, Maroc*
[4]*Nanophotonics Technology Center, Universitat Politècnica de València, Valencia, Spain*
[5] *Dept. of Physics, Universitat Autònoma de Barcelona, 08193 Bellaterra (Barcelona), Spain.*
[6] *ICREA - Institucio Catalana de Recerca i Estudis Avançats, 08010 Barcelona, Spain*

**Fabrication:** The 1D OM crystal structures were fabricated on standard silicon-on-insulator (SOI) samples of SOITEC wafers with a top silicon layer thickness of 220 nm (resistivity $\rho$ ~1-10 $\Omega\ cm^{-1}$, with a lightly p-doping of ~$10^{15}$ $cm^{-3}$) and a buried oxide layer thickness of 2 µm. The structure fabrication is based on an electron beam direct writing process performed on a coated 100 nm Poly(methyl methacrylate) (PMMA) resist film. The electron beam exposure, performed with a Raith150 tool, was optimized in order to reach the required dimensions employing an acceleration voltage of 10 KeV and an aperture size of 30 µm. The PMMA is a positive resist which implies that the negative of the patterns were exposed (window trenches and holes are exposed). The use of such positive resist allows getting well defined holes inside the waveguides as well as avoiding the use of a second lithography process in order to expose resist windows for selective silica removing. Nevertheless, the use of PMMA resist introduces fabrication imperfections (that could be improved by using negative resist as HSQ) in the interfaces between wider and narrower areas comprising the overall light waveguides.

After developing the PMMA resist using MIBK/IPA as developer, the resist patterns were transferred into the SOI samples employing an also optimized Inductively Coupled Plasma-Reactive Ion Etching process with fluoride gases (SF6/C4F8). Both gases are injected at the same time under typical plasma conditions for silicon etching (20 mT pressure, low bias voltage, and a SF6/C4F8 flow ratio of 1.5) which produce anisotropic etching with smooth sidewalls. Once the silicon is etched, a BHF silicon oxide etch (6:1) is finally used to remove the 2 micrometers BOX-silica layer in order to release the structures.

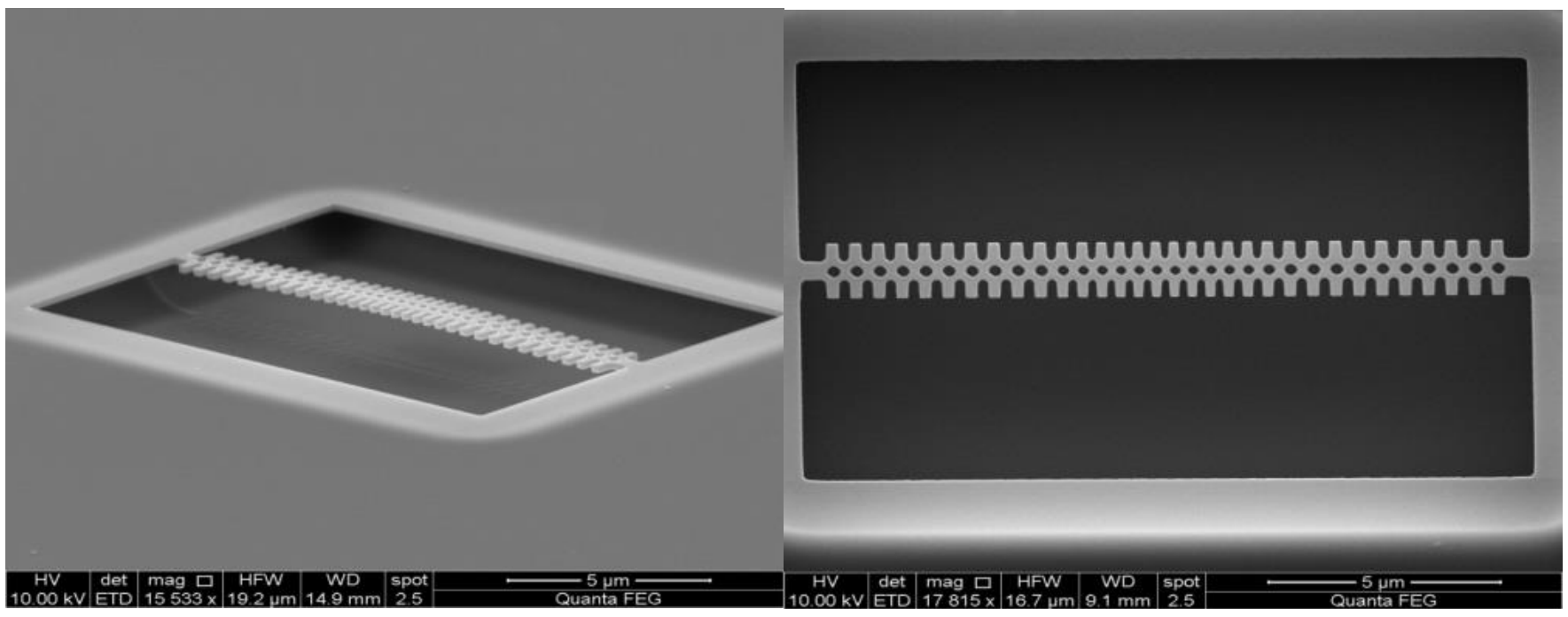

**Figure 1 SEM Images of typical 1D OM Crystals.**

**Fabrication Imperfections:** Several imperfections are noticeable in the SEM images (figure1). Rounding of the borders appears to be systematic, as the deformation of the holes into a rhombohedric form. The parallelicity of the stubs is affected by the rounding effect, and there are random variations in size and position. Dimensional and positions deviations are in the range of few nanometers.

We use the contour of the SEM images taken on the fabricated structures to model their optical and mechanical properties. With that procedure, the imperfections are taken into account naturally in the FEM calculations, and the optomechanical coupling rates and Q factors reported along the paper are already limited by fabrication imperfections.

**Experimental methods:** We use a tapered silica fibre loop to couple in and out light from the 1D OM crystal. We use the fabrication method described in reference [1]. The tapered fibre is fabricated using a microheater at 1170 º C while pulling the fibre extremes at a total rate of 40 μm/s. A laser set to the smallest wavelength used in the experiment transmits through the fibre during the fabrication process. Analysis of the time domain oscillations allow to monitor the transition to single mode of the tapered fibre zone. After tapering the fibre, we twist it to form a stable loop of around 30 μm of diameter (figure 2a), what limits the contact zone of the fibre with the sample. We place the fibre in contact to the 1D OM crystal, but also touching a border of the sample frame, as can be seen in figure 2. Looking at the optical transmission scanned fast (80 nm/s), the loading the fibre puts on the 1D OM crystal can be minimized.

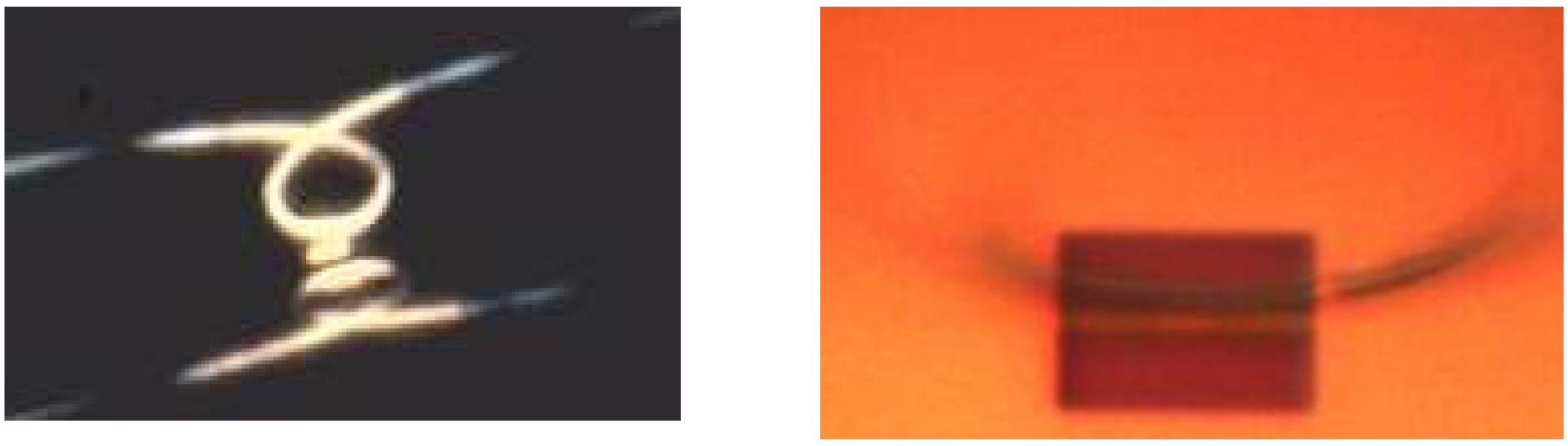

**Figure 2 Fibre to device adjustment a) Lateral image of a fibre loop with approximately 30 um diameter. b) Top microscope view of the sample-fiber adjustment.**

In figure 3, we show an scheme of the experimental setup used. Standard transmission measurements can be done by tuning the laser wavelegth in the IR (1460 to 1600 nm) with pm accuracy. For adjusting the fibre to device position we use a 100x microscope objective and a CCD cam to image from the top (figure 2 b). The fine tunnig of the position and a coarse adjustment of the polarization is performed by manually adjusting the positioning system with submicrometer precision, while observing in real time the transmission spectra. For that purpose a tunable laser (HP 8164 B system with a 81600 B laser source) is scanned over selected wavelength ranges at 80 nm/s scanning rate and transmission variations are collected with a slow response detector (HP 81634B).

The inestability of the scan laser, and the signal to source spontaneous emission (SSE) (>64 dB/nm) prevend its use for RF spectroscopy. Instead we use a more stable laser (Nettest Tunics-BT 3648 HE 1520, SSE<85 dB/nm). In addition, the fibre-sample system gets inestable and self-oscillate for high powers, what will be reported somewhere else. The power is kept low for all measurements presented in the present work (below 150 μW).

We perform RF spectroscopy with a 12 GHz photoreciever (New Focus 1544-B) in combination with a 20 GHz Signal Analyzer (Anritsu MS2830A). Amplification of the transduced signal is obtained by using an Erbium doped fibre amplifier (Keopsys CEFA-C-HG-SM-50-B201-FU-FU), what improves the signal to noise ratio and determines the preferred wavelength range for transducing the mechanical modes.

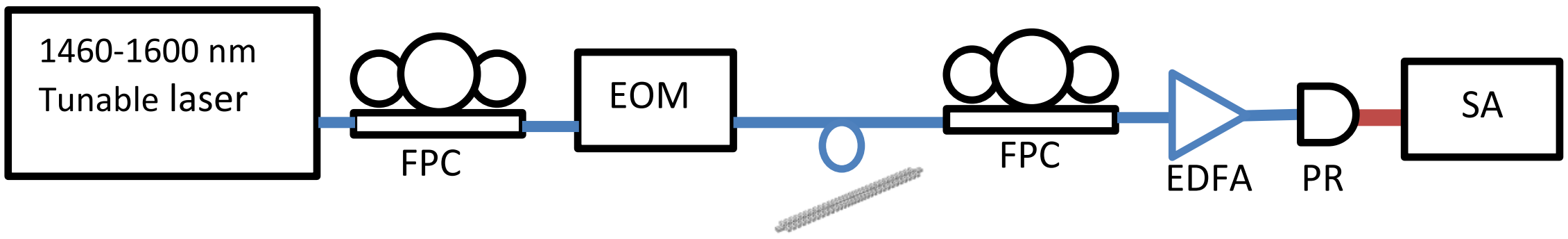

**Figure 3 Experimental setup. FPC Fibre Polarization control. EOM Electro Optical Modulator. EDFA Erbium doped fibre amplifier. PR Photoreceiver. SA Spectrum Analizer.**

**Optomechanical coupling calculation:** The calculation of $g$ is done using the formulations given by Chan et al.[2] for the photoelastic ($g_{\mathrm{PE}}$ ) and the moving interfaces ($g_{\mathrm{MI}}$) effects:

$$g_{\mathrm{PE}} = -\frac{\omega_0}{2}\frac{\langle E|\delta\varepsilon|E\rangle}{\int \mathbf{E}\cdot\mathbf{D}\, dV}\sqrt{\hbar/2m_{\mathrm{eff}}\omega_m}$$

$$g_{\mathrm{MI}} = -\frac{\omega_0}{2}\frac{\oint(\mathbf{Q}\cdot\hat{\mathbf{n}})\left(\Delta\varepsilon\mathbf{E}_{\parallel}^2 - \Delta\varepsilon^{-1}\mathbf{D}_{\perp}^2\right)dS}{\int \mathbf{E}\cdot\mathbf{D}\, dV}\sqrt{\hbar/2m_{\mathrm{eff}}\omega_m}$$

Where **Q** is the normalized displacement filed (max{|**Q**|}=1), $\hat{\mathbf{n}}$ is the outside normal to the boundary, **E** is the electric field and **D** the electric displacement field. $\varepsilon$ is the dielectric permittivity, $\Delta\varepsilon=\varepsilon_{silicon}- \varepsilon_{air}$ and $\Delta\varepsilon^{-1}=\varepsilon^{-1}{}_{silicon}- \varepsilon^{-1}{}_{air}$. $\boldsymbol{\delta\varepsilon} = -\varepsilon_0 n^4 p_{ijkl} S_{kl}$ , where $p_{ijkl}$ are the photoelastic tensor components, $n$ is the refractive index of silicon and $S_{kl}$ the strain tensor components. $\omega_0$ and $\Omega$ are the optical and acoustic frequencies respectively and $m_{\mathrm{eff}}$ is the effective motional mass related to the normalized acoustic displacement field by $m_{\mathrm{eff}} = \rho\int|\mathrm{Q}|dV$, where $\rho$ is the silicon density.

**Optomechanical coupling calibration:** We estimate the optomechanical vacuum coupling ($g$) strength from the experiments by calibrating the amplitude of the RF signal with a known reference as done and derived in reference [3]:

$$g^2 = \frac{1}{2\langle n\rangle}\frac{\beta^2\Omega_{mod}^2}{2\,RBW}\frac{\Gamma_m}{4}\frac{S_{\omega\omega}(\Omega_m)}{S_{\omega\omega}(\Omega_{mod})}$$

Where $S_{\omega\omega}(\Omega_m)$ and $S_{\omega\omega}(\Omega_{mod})$ are the peak powers measured in a SA of the OM transduced resonance ($\Omega_m$) and of the reference signal respectively ($\Omega_{mod}$). RBW: Resolution Bandwidth set on the SA. $\Gamma_m$ is the mechanical mode loss rate ($\Gamma_m = \Delta\Omega_m$ ) and a $\beta$ phase shift factor.

The spectral calibration is performed using the expression:

$$S_{\omega\omega}(\Omega_{mod}) = \frac{\beta^2\Omega_{mod}^2}{2\,RBW}$$

Our experimental layout is similar to the used in reference [4], the laser is tuned to the blue side of the optical resonance, transducing the movement as an amplitude modulation. We use the biased output of our photoreceiver to calibrate the phase factor $\beta$ of our doubled sided phase modulated reference. We use a Mach-Zender interferometer (Covega LN58S-FC) and we directly estimate the modulation ratio by comparing the DC level and the modulation amplitude of the reference with a fast oscilloscope (Tektronic TDS7404) after checking and correcting the frequency response of the full system, photoreceiver included, from DC to 10 GHz. The measurements are done using low power and checking that the spectral width is not affected by OM amplification.

We assume the occupation of the oscillator to be $n_{th}\sim\frac{k_B T_{th}}{\hbar\Omega_{\mathrm{mec}}}$ . The results from that calibration are compatible with the calculations presented in the paper. Using the fourth optical mode on a a device with a percentage reduction of 82% , we are capable of measuring a $g$ of 3.3 MHz ($g/2\pi$ = 525 KHz), as shown in Figure 4 (a), the highest in the set of devices fabricated. The modes confined in the phononic bandgap have limited optomechanical coupling with the actual design, as shown in Figure 4 (b).

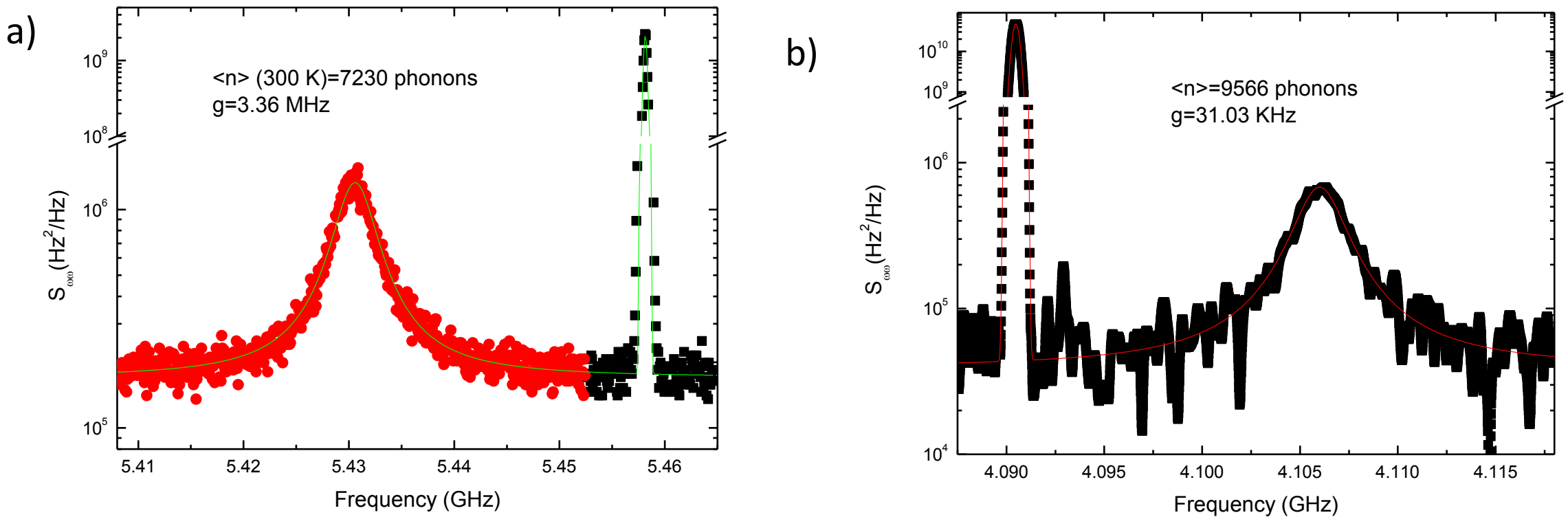

**Figure 4 Calibration of the noise floor and optomechanical coupling stimation for the mode with highest OM coupling ($g_{OM}$) (a) and the mode deeper in the complete phononic bandgap (b).**

We contrast the values obtained with this method with the optically-induced damping of the mechanical mode [5] method also used in optomechanical crystals. Red or blue detunig the laser from the cavity resonance ($\Delta=(\omega_0-\omega_l)=\pm\omega_m$) results in an optomechanically induced damping of $\gamma_{OM}(\pm\omega_m)=\pm 4G^2/\kappa$ where $|G|^2 = n_{cav}g$, $\kappa$ is the total optical loss rate and $n_{cav}$ is the amount of photons in the optical cavity. Nevertheless, the initial mechanical loss rate unaffected by the optomechanical backaction ($\gamma_i(T)$) is in the MHz range at room temperature, hiding small contributions. In addition, the power range available for the measurement is reduced in our experiment by thermo-optical effects that limit the access to the red detuned side of the optical resonance for $n_{cav}$ > 100. This avoids estimating the cooperativity slope measured as a function of power, but allows observing huge optomechanical amplification in the blue detuned case. Another source of noise is the wavelength stability of the laser cavity-detuning, normally solved by an stabilization electronic feedback our system lacks. As we place the tapered fibre in contact with the frame of the structure, we can only estimate an upper limit for $n_{cav}$ , what leads to underestimate $g$. All the aforementioned factors combined make this approach not practical to calibrate low $g$ and estimated big statistical errors for our experimental conditions. In any case, we have been capable of resolving both sidebands and estimating $g_0$ for the most optomechanicaly coupled phonon modes, as can be seen in Figure 5, and confirm the values obtained with the previous method.

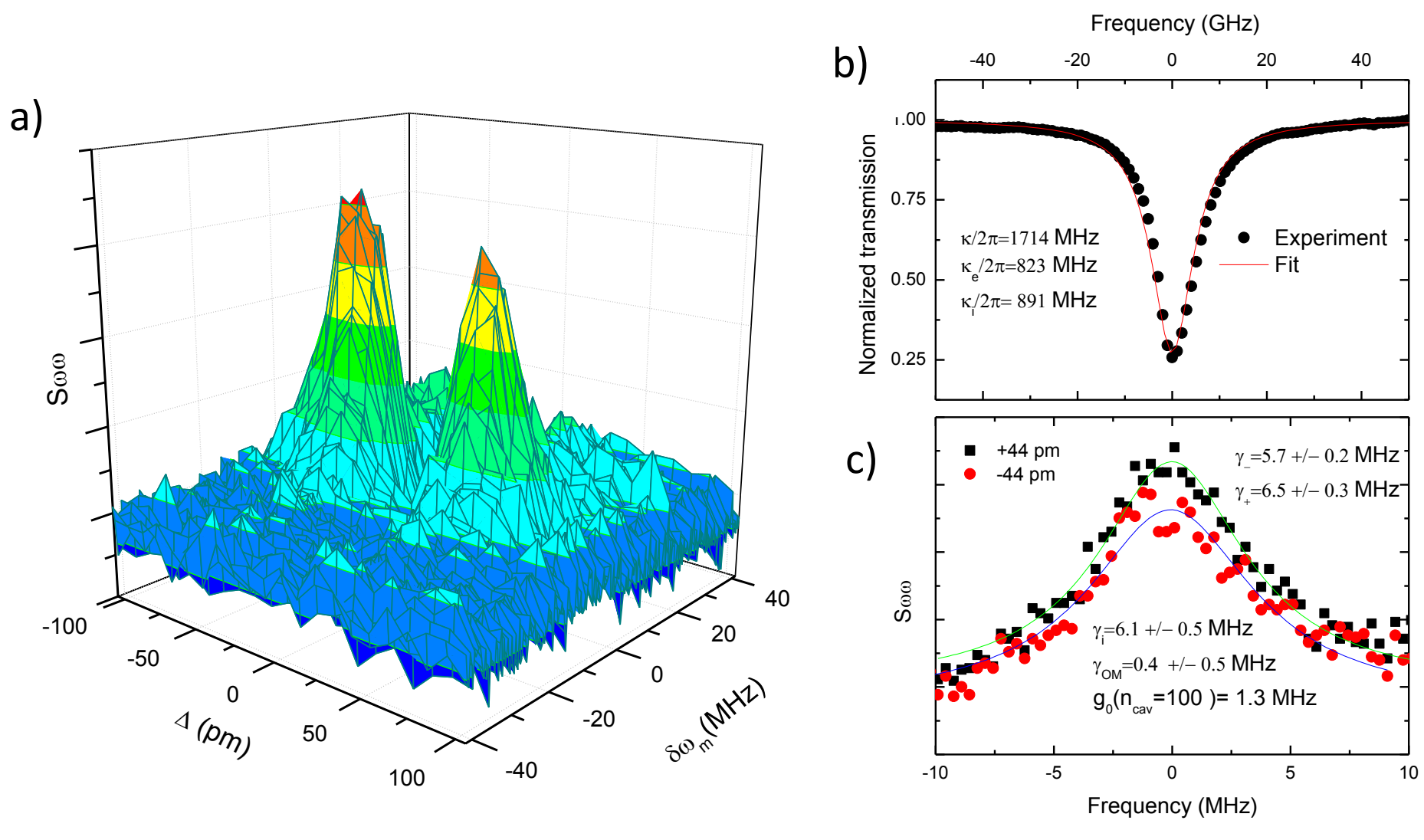

**Figure 5 Optomechanical coupling calibration with sideband resolved thermometry. (a) Transduction of the phonon mode as a function of the laser detuning (Δ). (b) Fit of the optical transmission (c) Comparison of the blue and red detuned RF spectra.**

## Selection of the phonon band by design:

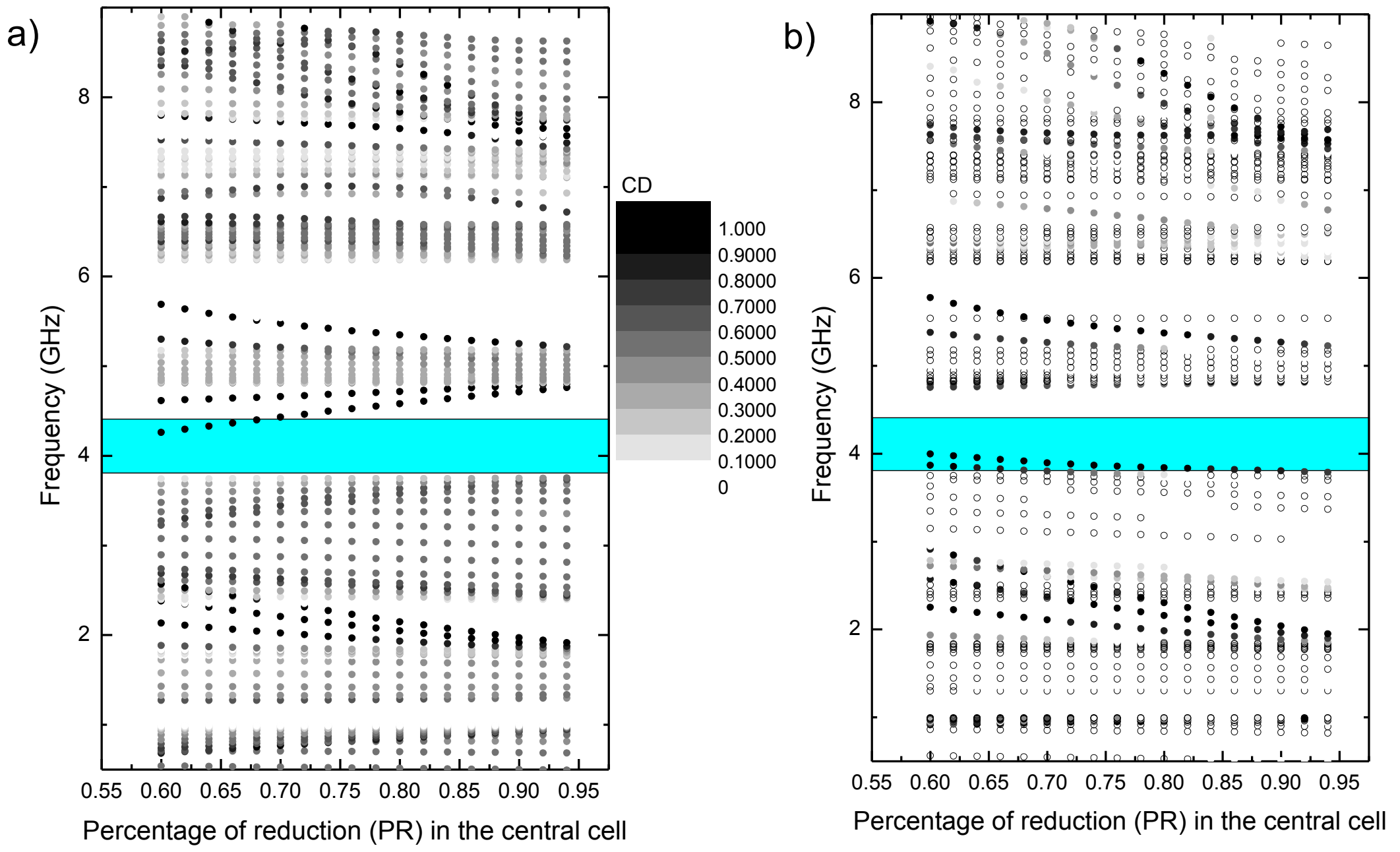

**Figure 6 Phonon modes evolution with the defect reduction. a) without the beam width variation b) with the beam width variation.**

The relative uncoupling of the photonic and phononic properties of the unit cell that we use allows slightly modifying the phononic properties of the defect without disturbing its optical performance. The decrease of the parameters defined in figure 1 of the paper leads to a systematic pull down of the phononic bands. To pull up the third symmetric-symmetric phononic band we have to include the beam width (w) increase that inverts the tendency. In Figure 6 we show a summary of the phonon frequencies found in the simulations when varying the design. We use an exemplary parabolic defect that allows us to describe the whole design with the percentage of reduction (PR) of the central cell with respect to the mirror cells. We calculate the degree of confinement (DC) of the phonon mode inside of the defect as the ratio of the displacement integrated inside the defect versus the displacement integrated in the whole structure (1 means fully confined). The points are grey scaled following the DC, what allows distinguishing easily between extended, clamped and confined phononic modes.
In Figure 6 it is possible to follow how a set of confined modes behaves around the complete phonon bandgap (3.8 to 4.3 GHz in these simulations) with the variation of PR. In a) the modes start from a high frequency band and decrease their frequency as PR decreases. Only for low PR a single band reaches frequencies below 4.5 GHz. In contrast, pulling the band up, as done in b) creates several confined modes set inside the bandgap for smaller values of PR.

**Loss mechanisms limiting the experimental Q factors:** Experimental measurements were performed at room temperature and at atmospheric pressure. The mechanical factor has a maximum value between 1500 and 2000 for phononic modes expanding almost a decade in frequency, showing little dependence on the frequency or the measured structure. The phonon decay time is limited by intrinsic phonon scattering mechanisms, like intrinsic phonon absorption[6,7] or thermo-elastic decay, as commonly accepted, even if we could not rule out extrinsic causes. We took special care to avoid touching the structure with the fibre, the fibre though, is in contact with the near frame.

It is well know that the thickness reduction in membranes affects the phonon dispersion, and as a consequence the phonon group velocity, density of states, specific heat capacity, electron-phonon and phonon-phonon interaction among others[8–13]. Also when the decrease in dimensions is in the order or smaller than the bulk phonon mean free path, diffusive boundary scattering can become dominant, what produces a dramatic decrease of the thermal conductivity[14]. The effective reduction of the phonon mean free path in nanostructured materials could produce a strong modification of the lifetime of phonons and as a consequence, a reduction of the mechanical Q factor.